\documentclass[smallcondensed,natbib]{svjour3}

\usepackage{latexsym}
\usepackage{microtype}
\usepackage{soul}
\usepackage{url}
\usepackage{graphicx} 
\usepackage{multirow}
\usepackage{xcolor}
\usepackage{tabularx}
\usepackage{booktabs}
\usepackage{diagbox}
\usepackage{hyperref}
\usepackage{subfig}
\usepackage{amsmath, ,amssymb}
\usepackage{todonotes}
\usepackage{dsfont}
\usepackage{etoolbox}

\newcommand{\senlabel}{\textsc{BERT-3S}}

\newcommand{\qalabel}{\textsc{Qa-DocRank}}

\newcommand{\qamodel}{\textsc{QA-Model}}

\newcommand{\clam}{\textsc{BERT based Rankers}}
\newcommand{\doclabel}{\textsc{Doc-Labelled}}
\newcommand{\bertcls}{\textsc{BERT-CLS}}
\newcommand{\msm}{\textsc{MS Marco}}
\newcommand{\trecdl}{\textsc{TREC-DL}}
\newcommand{\robust}{\textsc{Robust04}}
\newcommand{\core}{\textsc{Core17}}
\newcommand{\clueweb}{\textsc{ClueWeb09}}

\newcommand{\drmm}{\textsc{Drmm}}
\newcommand{\pacr}{\textsc{Pacrr}}
\newcommand{\mpcos}{\textsc{MatchPyramid}}
\newcommand{\pdrmm}{\textsc{PacrrDrmm}}

\newcommand{\mpara}[1]{\medskip\noindent{\bf #1}}

\begin{document}

\title{An In-depth Analysis of Passage-Level Label Transfer for Contextual Document Ranking\thanks{This paper is an extended work of our previous published paper "Distant Supervision in BERT-based Adhoc Document Retrieval". Koustav Rudra and Avishek Anand. In Proc. CIKM 2020. 2197-2200. In this paper, we extended this work over other datasets to understand the efficiency-efficacy trade-offs of different passage granularity.\\
This paper is accepted in Springer Information Retrieval Journal(\url{https://link.springer.com/journal/10791}).}}

\author{Koustav Rudra         \and
        Zeon Trevor Fernando \and
        Avishek Anand
}


\institute{Koustav Rudra \at
              Indian Institute of Technology Kharagpur, India \\
              \email{krudra@cai.iitkgp.ac.in}           
           \and
           Zeon Trevor Fernando \at
             Immobilescout, Berlin, Germany \\
             \email{zeon.trevor@gmail.com}
           \and
           Avishek Anand \at
             Delft University of Technology, Netherlands \\
             \email{Avishek.Anand@tudelft.nl}
}

\maketitle

\begin{abstract}
Pre-trained contextual language models such as BERT, GPT, and XLnet work quite well for document retrieval tasks. Such models are fine-tuned based on the query-document/query-passage level relevance labels to capture the ranking signals. However, the documents are longer than the passages and such document ranking models suffer from the token limitation (512) of BERT. Researchers proposed ranking strategies that either truncate the documents beyond the token limit or chunk the documents into units that can fit into the BERT. In the later case, the relevance labels are either directly transferred from the original query-document pair or learned through some external model. In this paper, we conduct a detailed study of the design decisions about splitting and label transfer on retrieval effectiveness and efficiency. We find that direct transfer of relevance labels from documents to passages introduces \textit{label noise} that strongly affects retrieval effectiveness for large training datasets. We also find that query processing times are adversely affected by fine-grained splitting schemes. As a remedy, we propose a careful passage level labelling scheme using weak supervision that delivers improved performance (3-14\% in terms of nDCG score) over most of the recently proposed models for ad-hoc retrieval while maintaining manageable computational complexity on four diverse document retrieval datasets.

\end{abstract}

\section{Introduction}
\label{sec:intro}

In web search and information retrieval, document retrieval is a standard task whether the objective is to rank the documents with respect to a query such that most relevant documents appear on top of the list. Neural ranking approaches~\citep{matchpyramid16,pacrr17,pacrr_drmm_18} show better performance over term-matching based strategies~\citep{strohman_indri_2005}. Recent, pretrained language model based methods~\citep{dai_sigir_2019,macavaney2019contextualized:bertir:mpi,lin_emnlp_2019,rudra_2020_distant,li2020parade} show significant improvement because they are able to understand the intent of a short query through contextual interaction with the documents. However, the major issue with this language models is the limitation of input token. Even if some recent models (e.g., xlnet) support large input lengths, the problem of gradient vanishing is there.

\begin{table}[tb]
    \footnotesize
    \centering
    \begin{tabularx}{\textwidth}{X}
    \toprule
        \textbf{Passage 1:} \\
        \cmidrule(lr){1-1}
         \textit{Eventually \colorbox{pink}{Parkinson's surges forward}, leaving advancing dysfunction and death in its wake. For me, that -- never mind my career, or my new marriage or my dreams of having children -- is the future. The millions of other Americans \colorbox{pink}{afflicted with Parkinson's}, diabetes and the other diseases have their own stories of unrealized dreams; of watching their bodies fail them, and being unable to do anything to stop it.} \\
        \cmidrule(lr){1-1}
        \textbf{Passage 2:}\\
        \cmidrule(lr){1-1}
        \textit{`` The bill, which could free such funding from political intervention, faces its first vote this month. And when it does, a congressional "pro-life" force is expected to attempt to reduce the discussion to the anti-abortion rhetoric that was used to justify the presidential moratorium. ''}
        \\
        \bottomrule
    \end{tabularx}
    \caption{{ Two sample passages for the query \textbf{``Parkinson's disease''} taken from the document marked as relevant by human annotators. However, the first passage is relevant to the query while the second one is not.}}
    \label{tab:sample_example}
\end{table}

The most common design choice used in retrieval and other NLP tasks when dealing with long documents is truncating or considering a limited part of the document~\citep{macavaney2019contextualized:bertir:mpi}. 
However, truncation leads to undesirable information loss.
Consequently recent approaches have tried dividing the document into passages~\citep{dai_sigir_2019}, or sentences~\citep{lin_emnlp_2019}. 
Specifically,~\citet{dai_sigir_2019} transfers the document level relevance label to each of the passages.
However, not all passages in a document are relevant. Table~\ref{tab:sample_example} shows examples of relevant and non-relevant passages from the same document for the query `\textbf{Parkinson's disease}'. 

Another dimension that makes machine learning for ad-hoc retrieval a challenging task is due to \textit{label sparsity} and \textit{label noise}.
Firstly, the training labels are sparse because not all relevant query-document pairs are labelled due to large document collection sizes and exposure bias effects due to the ranking of documents~\citep{craswell2008experimental}. 
Secondly, gathering user assessments for documents (relevant or not) given under specified queries using implicit feedback~\citep{white2002comparing,kelly2003implicit} techniques adds label noise.
These limitations have resulted in two types of datasets being available to the IR community -- \textit{mostly labelled} small dataset of queries like TREC Robust data or partially labelled large dataset of queries derived from implicit feedback (i.e. query logs) like \trecdl{} dataset~\citep{msmarco_trec_2019}.
In this context, passage label assignments as in~\citet{dai_sigir_2019},  derived from document-level assessments, are yet another source of label noise. 

In this paper, we follow a simple premise. If we can only consider a subset of relevant passages for training, then we can significantly reduce the noise in the label assignment to passages.
Towards this, we build on the recent finding by~\citet{lin_emnlp_2019} who show that document retrieval performance can be improved by using a model trained on retrieval tasks that do not exhibit input length limitation problems. 
Specifically, unlike~\citet{lin_emnlp_2019} who use an external model only during inference, we use an external passage ranking model for QA tasks (\qamodel) to label relevant passages prior to training.
In our examples in Table~\ref{tab:sample_example}, the \qamodel{} marks the second passage as irrelevant to the query as desired.
Apart from potential noise reduction in passages, such a simple labelling scheme has implications in improving training and inference efficiency as well.
In sum, we ask the following research questions and summarize our key findings on the effectiveness of passage-level label transfer for document retrieval:

\begin{itemize}
    \item
    What impact do large training datasets have on different labelling strategies of contextual ranking models?
    
    \item
    What impact do models trained on different collections have on the ranking performance?
    
    \item
    What is the impact of transfer-based contextual ranking models on the efficiency of {\em training} and {\em inference}?
    
\end{itemize}

Note that, our distant supervision based architecture was proposed in~\citet{rudra_2020_distant}. In this paper, we have performed detailed set of experiments to understand the robustness and efficiency of our proposed approach.
\begin{enumerate}
    \item Earlier, we tested our approach only over a small fraction of \trecdl{} training and development set. In the current work, we validate the performance of our proposed approach over the recent test set of \trecdl. The performance over the entire training set (367K queries) and variation in performance over different training sizes are also checked. Side by side, we also validate the efficacy of \qamodel{} on three more datasets \robust, \core, and \clueweb.
    
    \item In our previous version, the transfer model was trained on MSMARCO passage levels and applied over \trecdl{} to judge the query-passage relevance. However, both the datasets come from the same distribution and we don't have such passage level counterpart for other standard document retrieval datasets such as \robust, \core, and \clueweb. In this paper, we apply the transfer model trained on MSMARCO over different document retrieval datasets and surprisingly it performs quite well for other datasets. It gives a signal that transfer knowledge works efficiently in document reranking.
    
    \item There is a significant dependency between the document chunking procedure and document ranking. In this paper, we have shown the influence of document chunking on the performance of different transfer models over various types of datasets. We also highlight the necessity of efficient label transfer in maintaining the robustness of the ranker models.
    
    \item In this work, we also analyze the efficiency of different BERT based models in terms of model training and inference time. Apart from that, we also explore the role of zero shot learning in document retrieval. We observe that recent contextual models may be directly applied over a new dataset for ranking and competitive performance may be achieved based on the selection of appropriate dataset(Section~\ref{sec:zero-shot}).
    
    \item We have uploaded our code in Github (\url{https://github.com/krudra/QADocRank}).
\end{enumerate}

\noindent\textbf{Key Takeaways} -- We conduct extensive experiments on four TREC datasets of different collection and label properties. We show that our distantly supervised retrieval model (\qalabel) is highly sample efficient as compared to document level label transfer. In fact, distantly supervised training of BERT-based models outperforms most of the existing baseline models.
We also find that cautious cross-domain knowledge transfer from a QA passage ranking model helps balance between {\bf retrieval performance} and {\bf computational complexity}.

\section{Related Work}
\label{sec:related}

Ad-hoc retrieval is a classical task in information retrieval where there is precedence of classical probabilistic models based on query likelihood~\citep{lavrenko2017relevance} and BM25~\citep{robertson_bm25_2009} proving hard to beat. In recent times, neural models have played a significant role in ad-hoc document retrieval and reranking. Researchers not only focused on the performance issue but also on the efficiency issues of the models. In this section, we give a brief description of different categories of ranking models.

\textbf{Neural Models:} Neural models bring significant changes in modeling queries and documents. They help in getting the semantic representations~\citep{dssm13,Shen2014a,Shen2014b}, positional information~\citep{pacrr17,co_pacrr_wsdm18,pacrr_drmm_18}, local query-document interactions~\citep{Guo2016,matchpyramid16,KNRM17,Nie_ictir18,Nie_sigir_2018} or a combination of both~\citep{Mitra2017a}. Broadly, there are representation and interaction based models that explore query and document representations and interactions between them. Representation models present queries and documents in low dimensional latent feature space by passing them through deep neural networks and then computing the similarity between the vectors. Huang~et~al~\citep{dssm13} passed queries and documents through simple feed forward networks to get the semantic representations and measure the similarity score based on those vectors. Shen~et~al~\citep{Shen2014b} used CNN instead of feed forward network to capture the local context window. CNN is also used in many other representation based neural ranking models that rely on semantic representation of queries and documents~\citep{hu_cnn_2014,Shen2014a,qiu_cnn_qa_2015}.
Another line of work focuses on the word sequences in the queries and documents and they represent them using sequence aware models such as RNN or LSTM~\citep{hochreiter1997long}. LSTM is used to learn the vector representations of queries and documents and finally measure the similarities between those vectors using cosine similarities~\citep{mueller_rnn_doc_2016,palangi_lstm_doc_2016}. Later on, Wan~et~al~\citep{wan_lstm_doc_2016} proposed Bi-LSTM based representation of queries and documents, and the final similarity is measured through a neural layer.  

In the representation based models, query and document feature vectors are learned independently and their interaction is deferred up to the last stage. Hence, most of the important matching signals get missed and it affects the performance of the document ranker. Hence, researchers proposed interaction based models over representation ones. Guo~et~al~\citep{Guo2016} proposed a Deep Relevance Matching Model (DRMM) that first learns an interaction matrix between query and document using embeddings of query and document tokens. From this matrix, DRMM learns histogram based matching patterns to predict the relevance of query-document pairs. DRMM relies on hard assignment and it poses a problem for backpropagation. Hence, Xiong~et~al~\citep{KNRM17} proposed a kernel pooling based soft matching approach to overcome this limitation. Several interaction based approaches such as Hierarchical Neural matching model (HiNT)~\citep{Fan_hint_2018}, aNMM~\citep{yang_anmm_2016}, MatchPyramid~\citep{matchpyramid16}, DeepRank~\citep{Pang_deeprank_2017}, Position-Aware Convolutional Recurrent Relevance (PACRR)~\citep{pacrr17} rely on interaction matrix and similarity measures like cosine similarity, dot product, etc.

A CNN is used in many interaction based models~\citep{pacrr17,dai_ConvKNRM_18,Nie_sigir_2018,pacrr_drmm_18,tang_2019_deeptilebars}. In general, such models use different size kernels (1D, 2D) in multiple layers of CNN and finally predict the query-document level relevance score using some MLP at the final layer. Dai~et~al~\citep{dai_ConvKNRM_18} extends the idea of KNRM~\citep{KNRM17} in their Conv-KNRM model that uses CNN filters to compose n-grams from query and documents and the embeddings of such n-grams are used to learn the similarity between query-document pairs. PACRR-DRMM~\citep{pacrr_drmm_18} consider the modeling benefits of both PACRR~\citep{pacrr17}  and DRMM~\citep{Guo2016} i.e., it learns document aware query token encoding in the place of histogram in DRMM.

Along with CNN, sequential neural models (RNN, GRU, LSTM) also play a key role in interaction based reranking approaches. Several approaches used LSTM based modeling of queries and documents~\citep{fan_sigir_2018}. Wan~et~al~\citep{wan_lstm_doc_2016} proposed Match-SRNN based on GRU to accumulate matching signals. In a similar line, MatchPyramid~\citep{matchpyramid16} and DeepRank~\citep{Pang_deeprank_2017} fed the interaction matrix between the query and document to a GRU to learn the final feature vector. Some models such as DUET~\citep{Mitra2017a} combined the benefit of both representation (distributed model) and interaction (local model) based networks to achieve better reranking performance. 

~\\
\textbf{Deep Contextualized Autoregressive Neural Model based Rankers:}
Recently introduced pre-trained language models such as ELMO~\citep{peters_2018_elmo}, GPT-2~\citep{radford2019language}, SentenceBERT~\citep{reimers_2019_sentencebert}, and BERT~\citep{devlin_bert_2018} show promising improvement in different NLP tasks. Such models are trained on huge volumes of unlabelled data. Such contextual models, e.g., BERT, have proven to be superior in the document reranking task than the above neural models.
The sentence classification task of BERT is extensively used in BERT based document retrieval techniques~\citep{dai_sigir_2019,nogueira2019multistage,yang2019simple:bertir:lin, wu_passage_level_relevance_2020}.
Previous models addressed BERT's fixed input restriction either by sentence-wise labelling~\citep{lin_emnlp_2019} or passage-level labelling~\citep{dai_sigir_2019,wu_passage_webconf_2020}. \citet{dai_sigir_2019}. \citet{dai_sigir_2019} split documents into passages, and obtain passage level relevance score by fine-tuning the BERT model (\doclabel).

On the other hand,\senlabel~\citep{lin_emnlp_2019} is a cross-domain knowledge transfer based modeling approach. It splits documents into passages and computes the score of each query-sentence pair using a secondary model trained on the MSMARCO passage and Microblogging retrieval dataset. Finally, it computes the relevance score of a query-document pair by interpolation between a sparse index based score (BM25/QL score) and the semantic score of the top three sentences learned via the transfer model.
\doclabel{} approach considers all passages of a relevant document as relevant and this introduces label noise. \senlabel{} method does sentence-level knowledge transfer and takes a large inference time. This approach performs contextual modeling of documents and provides a useful upper bound on performance.

Subsequently, MacAvaney~et~al~\citep{macavaney2019contextualized:bertir:mpi} combined the power of BERT (contextual representations) and interaction based models such as Conv-KNRM~\citep{dai_ConvKNRM_18}, KNRM~\citep{KNRM17} to improve the performance of ranking models. Recently, Li~et~al~\citep{li2020parade} proposed an end-to-end PARADE method to overcome the limitation of independent inference of passages and predict a document's relevance by aggregating passage representations. Hofstatter~et~al~\citep{hofstatter2020local,hofstatter2020interpretable} proposed local self-attention strategies to extract information from long text. This transformer-kernel based pooling strategy becomes helpful to overcome the fixed length token limitation of BERT. Side by side, this kernel based strategy consumes less amount of parameters than BERT models. 

So far, all the BERT-based approaches jointly modeled query-document sequences (cross-attention). This incurs huge computational costs, especially in inference time where we have to rerank around hundred to thousand documents per query. Such large transformer based models show better performance at the cost of orders of magnitude longer inference time~\citep{hofstatter2019let,macavaney2019contextualized:bertir:mpi}.
To overcome this restriction, researchers also proposed independent modeling of queries and documents.
Dual encoder architecture is a strategy that encodes query and document independently of each other~\citep{lee2019latent,ahmad2019reqa,chang2020pre,karpukhin2020dense,khattab2020colbert,hofstatter2020interpretable}.
This shows promising results both in terms of performance and inference cost. The BERT model of the document arm is only fine-tuned during the training phase but froze in the testing phase.
Query-independent latent document representations~\citep{luan2020sparse} make the precise matching of terms and concepts difficult and therefore explicit term matching methods are also combined along with latent representations~\citep{Nalisnick:2016,Mitra2016a}.
Xiong~et~al~\citep{xiong2020approximate} have recently established that the training data distribution may have a significant influence on the performance of dual encoder models under the full retrieval setting. 
Tilde~\citep{zhuang2021fast} and Tildev2~\citep{zhuang2021tilde} proposed a deep query and document likelihood based model instead of a query encoder to improve the ranking efficiency. The SpaDE~\citep{choi2022spade} model improves the ranking efficiency by using simplified query representations and a dual document encoder containing term weighting and term expansion components. Other approaches also tried to improve the ranking efficiency by compressing document representations~\citep{cohen2022sdr} and removing unnecessary word representations (COLBERTER)~\citep{hofstatter2022introducing}. Further, researchers also explored \textit{hybrid models} where they interpolate between the scores of sparse and dense retrieval models. There exist several models in this line such as CLEAR~\citep{gao2020complement}, COIL~\citep{gao2021coil}, COILCR~\citep{fan2023coilcr}. Anand~et~al~\citep{anand2023data} proposed a data augmentation based robust document retrieval framework. 
Leonhardt~et~al~\citep{leonhardt_explainable_ranking,leonhardt_ff_www_2022,leonhardt_ff_tois_2023} focused on the interpretability and efficiency of the document retrieval models.
\textit{In this paper, we particularly focus on BERT based reranking approaches that jointly model query and document sequences using cross attention approach. Our objective is to explore the trade-off between {\it effective transfer} and {\it efficient transfer} rather than new architectural improvements as in~\citep{macavaney2019contextualized:bertir:mpi,li2020parade,hofstatter2020interpretable,choi2022spade,fan2023coilcr,leonhardt_ff_tois_2023}}. However, we believe our study could be extended to other kinds of reranking models such as dual encoder, hybrid models.


~\\
\textbf{Weak supervision:} Another line of work tried to train neural ranking architectures using large-scale weak or noisy labels~\citep{Dehghani_sigir17,dehghani2018fidelityweighted,zhang_weak_wbconf_2020}. The teacher-student paradigm~\citep{hinton2015distilling,xiao2015learning} is also used to infer better labels from noisy labels. These labels are further used to supervise the network training ~\citep{sukhbaatar2014training,veit2017learning}. 
Although similar in spirit to our approach, our work falls in the intersection of transfer learning and weak supervision in that we judiciously select a subset of training instances from the original training instances (passage-query pairs) using a model trained on another task.

\newcommand{\cls}{\texttt{[CLS]}}
\newcommand{\sep}{\texttt{[SEP]}}
\newcommand{\bert}{\texttt{BERT}}
\newcommand{\q}{\mathbf{q}}
\newcommand{\doc}{\mathbf{d}}

\begin{figure}[tb]
\centering
    \includegraphics[width=0.9\textwidth]{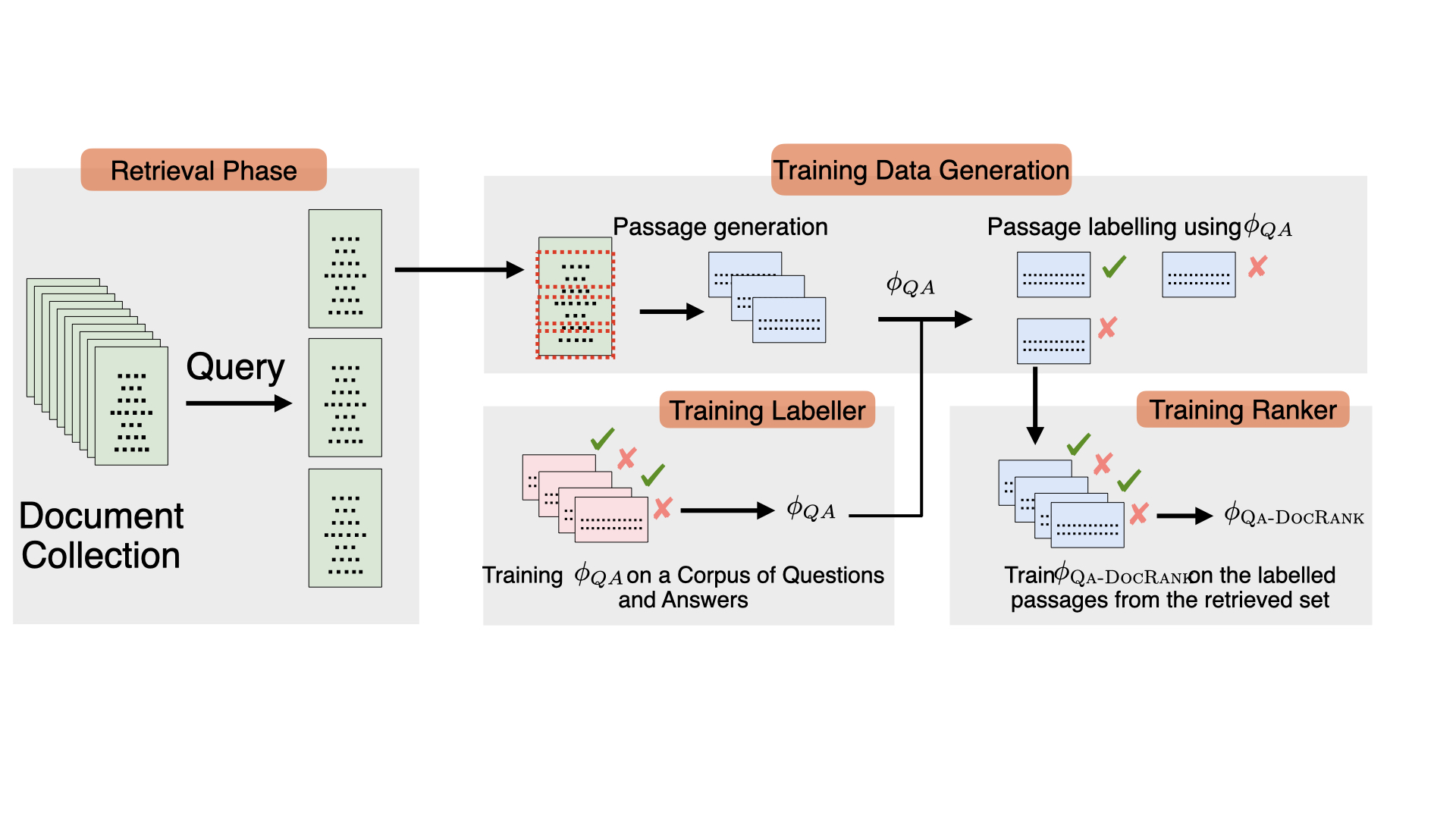}
    \caption{{ Training $\phi_{\qalabel}$}}
    \label{fig:qa-docrank}
\end{figure}
\section{Document Ranking with Passage Level Label Transfer}
\label{sec:method}

Typical approaches to the ad-hoc retrieval problem follows telescoping setup~\cite{Matveeva06} and consists of two main stages. 
First, in a retrieval phase, a small number (for example, a thousand) of possibly relevant documents to a given query are retrieved from a large corpus of documents by a standard retrieval model such as BM25 or QLM~\cite{Lavrenko_2001}. In the second stage, each of these retrieved documents is scored and re-ranked by a more computationally-intensive method.
Our focus is in the ranking problem in the second stage using contextual ranking models based on \bert{} introduced in~\cite{devlin_bert_2018}.

\subsection{Limitation of \bert}
\label{sec:bert_limitation}
\cite{nogueira_prr_2019} were the first to show the effectiveness of contextual representations using \bert{} for the passage reranking task for QA on the \msm{} dataset. They proposed \qamodel{} for the passage reranking task.
However, the maximum token size for \bert{} is 512. This fits quite well for sentence pair task or question-passage task. Documents are longer than sentences and passages and it is difficult to fit them to \bert{} model. This poses a real challenge to the query-document ranking task.
In the following sections, we present our approach of handling long documents in \bert{}. The overall training process is outlined in Figure~\ref{fig:qa-docrank}.

\subsection{Passage Generation and Labeling}
\label{sec:passage_labeling}
\mpara{Passage Generation.} 
We follow the basic framework of~\cite{dai_sigir_2019} in dealing with long documents in that documents are chunked into passages of fixed size. Specifically, we follow the approach proposed by~\cite{fan_sigir_2018} for passage generation. 
That is, we first prepend the title to the document text.
We then split each document into passages of length $100$ (white-space tokens). If the last sentence of a passage crosses the word boundary of $100$, we also take the remaining part of that sentence into the current passage.

\mpara{Passage Labeling.} Unlike~\cite{dai_sigir_2019} that indiscriminately transfers document relevance labels to all its passages, in this paper we follow an alternate labelling scheme to selectively label passages of a relevant document. 
Our idea is to use an external model that is trained on a different (yet related) task of finding relevant passages given a query as a labeler for our generated passages.
Towards this, we choose the model proposed by~\cite{nogueira_prr_2019} for passage reranking task and refer to this model as \qamodel{} :

\begin{align}
\label{eqn:qamodel}
    \phi_{QA}: (\q,\mathbf{p}) \xrightarrow{} \: \{  \mathtt{relevant},\mathtt{\neg relevant} \}.
\end{align} 

This model makes use of \bert{} architecture. In this part, we introduce the \bert{} architecture first and then show the working procedure of \qamodel. Nogueira and Cho~\citep{nogueira_prr_2019} used BERT's sentence-pair model to get the relevance score of each query-passage pair. BERT is trained on an unsupervised language modeling (LM) task on English Wikipedia and Book corpus datasets. Two different LM tasks (\textit{Masked LM} and \textit{Next Sentence Prediction}) are chosen to optimize the BERT model. In \textit{Masked LM}, some words are randomly chosen and they are replaced either with [MASK] token or a random word. The goal of the \textit{Masked LM} task is to predict the masked word correctly. Given the two sentences, the objective of the \textit{Next Sentence Prediction} is to decide whether two sentences in a paragraph appear next to each other. BERT learns to represent sentences in the process of learning the above mentioned two tasks over a large text corpora. Thats why pre-trained BERT contains lots of parameters, e.g., $BERT_base$ contains around 110M parameters. Pre-trained BERT can be fine-tuned for several other NLP tasks. Nogueira and Cho~\citep{nogueira_prr_2019} used pre-trained BERT model to get the relevance of a query-passage pair. In this process, all the parameters of BERT are also fine-tuned in an end-to-end manner for the query-passage relevance detection task. BERT can be viewed as a combination of multilayer Transformer network~\citep{vaswani_2017_attention}.

\begin{figure}[tb]
\centering
    \includegraphics[width=0.9\textwidth]{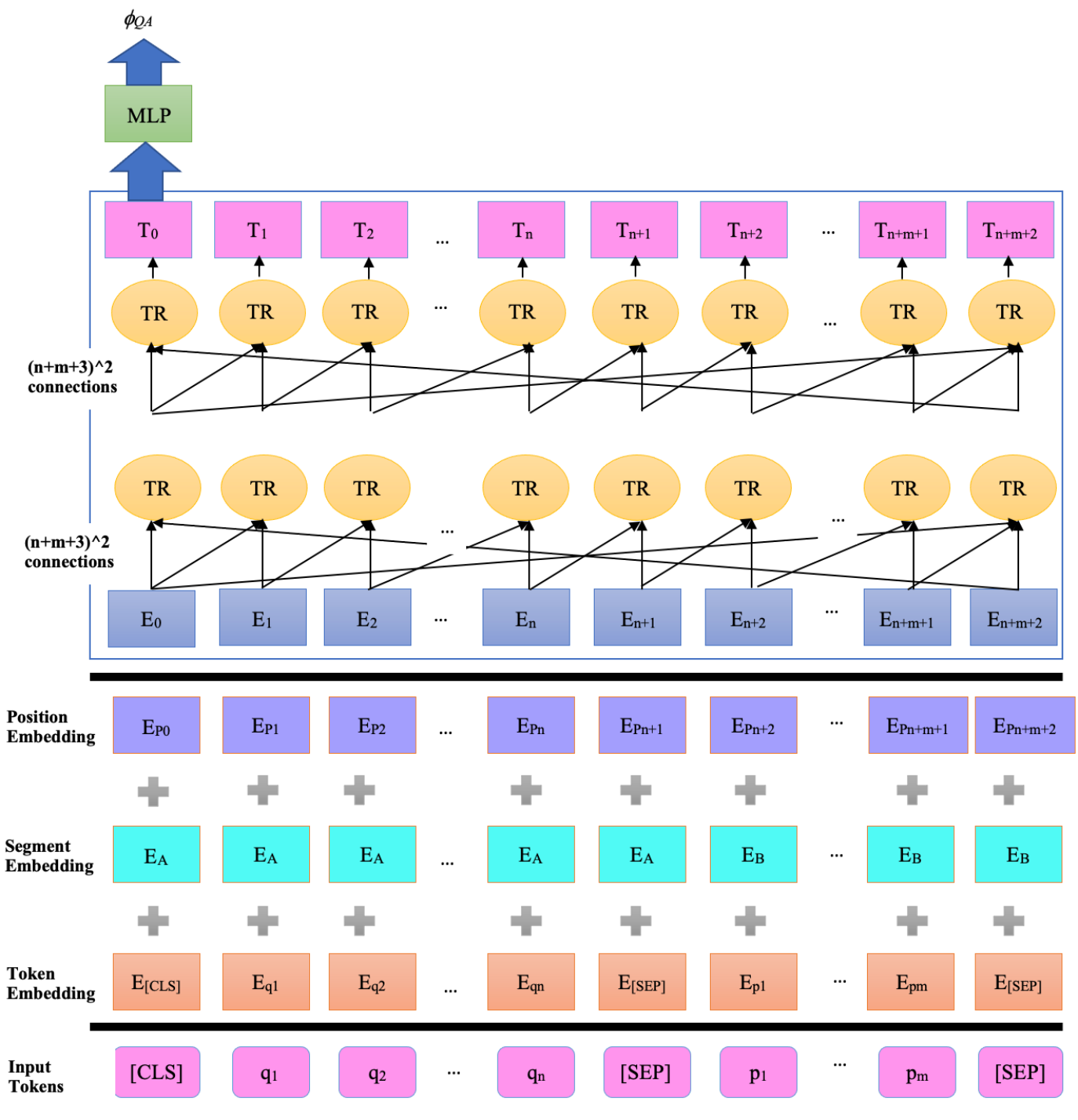}
    \caption{{${\qamodel(\phi_{QA})}$ model architecture. The BERT model takes a query q = ${q_{1}, q_{2}, ..., q_{n}}$ and passage p = ${p_{1}, p_{2}, ..., p_{m}}$ of length $n$ and $m$ respectively. This is passed through several transformer (TR) layers and finally representation of the $CLS$ token is passed through a feed forward layer to predict the relevance score of the query-passage pair.}}
    \label{fig:qamodel}
\end{figure}

Technically, this is realized by forming an input to \bert{} of the form 
[ \cls{}, $\q$, \sep{}, $\mathbf{p}$, \sep{} ] and padding each sequence in a mini-batch to the maximum length (typically 512 tokens) in the batch. 
The final hidden state corresponding to the \cls{} token in the model is fed to a single layer neural network whose output represents the probability that passage is relevant to the query $\q$. Fig.~\ref{fig:qamodel} depicts the framework.

We use the trained \qamodel{} (Equation~\ref{eqn:qamodel}) to obtain relevance labels for query-passage pairs derived from the initial retrieved set of documents. 
Specifically, we label a passage $\mathbf{p} \in \doc$ of a \textbf{relevant document} as $\mathtt{relevant}$ if $\phi_{QA}(\q,\mathbf{p}) = \mathtt{relevant}$.

\subsection{Proposed Document Ranking Model}
\label{sec:qabasedmodels}

We now detail our training and inference procedure on the newly labelled query-passage pairs.

\mpara{Training.}
After obtaining the passage labels, we now finetune another BERT model on these query-passage pairs following the approach proposed by~\citet{nogueira_prr_2019}. 
That, using the pre-trained BERT base uncased model~\citep{devlin_bert_2018}, we fine-tune the model end to end using the passage level relevance signal. 
We randomly sample same number of non-relevant query-passage pairs as relevant ones because the number of non-relevant pairs are way larger than the relevant ones.
In some sense our setup resembles a teacher-student training paradigm where our \qamodel{} works as a \textit{teacher} to determine relevant passages in a document to assist the \textit{student}, here \qalabel{}, in the training process.

\mpara{Inference.}
Finally the trained \qalabel{} model is applied over the test set to predict the relevance labels of query-passage pairs and these scores are \textit{aggregated} to get the final score of the corresponding query-document pair.
\textit{Note that, this \qamodel{} based passage-level judgements are only applied to training and validation sets. For the test set, documents are ranked based on different aggregation methods  applied over passage level scores.}
Towards this, we adopt four different aggregation strategies as proposed by ~\citet{dai_sigir_2019}. Apart from that, we use two position aware passage score aggregation strategies. The aggregation functions are given below:

\begin{enumerate}
    \item {\bf FirstP:} Score of the first passage
    \item {\bf MaxP:} Score of the best passage
    \item {\bf SumP:} Sum of all passage scores
    \item {\bf AvgP:} Average of all passage scores
    \item {\bf DecaySumP:} Instead of giving equal weight to all the passages when summing up the scores, the passage weights are multiplied by the inverse of their position in the document.
    \begin{equation}
        \psi(\q,\doc) = \underset{\mathbf{p_{i}} \in \doc}{\sum_{i=1}^{m}}(\phi_{\qalabel}(\q,\mathbf{p_{i}})*\frac{1}{i})
    \end{equation}
    \item {\bf DecayAvgP:} Similar to DecaySumP, but the the total score is normalized by the number of passages.
    \begin{equation}
        \psi(\q,\doc) = \frac{\underset{\mathbf{p_{i}} \in \doc}{\sum_{i=1}^{m}}(\phi_{\qalabel}(\q,\mathbf{p_{i}})*\frac{1}{i})}{m}
    \end{equation}
\end{enumerate}

\begin{align}
\label{eqn:qa_partial_model_inference}
\psi(\q,\doc) = \underset{\mathbf{p} \in \doc}{\mathtt{Aggr}}(\phi_{\qalabel}(\q,\mathbf{p}))
\end{align}

In our experiments we refer to the ranking based on scoring after aggregation (Equation~\ref{eqn:qa_partial_model_inference}) as \qalabel{}.

\section{Experimental Evaluation}
\label{sec:experiment}

In this section, we experimentally evaluate the efficiency of our proposed approach \qalabel. 
We begin by describing our baselines, experimental setup, and evaluation procedure in this section.

\subsection{Baselines and Competitors}
The first-stage retrieval model, the query likelihood model, is also considered as a ranking baseline~\citep{Lavrenko_2001}. 
Our competitors are the following Non-contextual and contextual rankers.

\mpara{Non Contextual Neural Models.} We then compare against non-contextual neural neural ranking models \pdrmm~\citep{pacrr_drmm_18} that combines the modelling of \pacr~\citep{pacrr17} and aggregation of \drmm~\citep{Guo2016}. We also tried other non-contextual neural models like \mpcos~\citep{matchpyramid16}, \drmm{}, and \pacr{} but \pdrmm{} consistently outperforms them. Hence, due to space constraints we use \pdrmm{} as a representative non-contextual neural model and  skip other results.

\mpara{Contextual Ranking Models.} We consider following contextual ranking models.
\begin{enumerate}
 \item \textbf{\doclabel}\citep{dai_sigir_2019}: Baseline from~\citet{dai_sigir_2019} where relevance labels are transferred from the document level to passage level.
 
 \item \textbf{\bertcls}\citep{devlin_bert_2018}: The BERT model is fine-tuned with document level supervision. This is truncation based approach~\cite{macavaney2019contextualized:bertir:mpi} where content beyond 512 tokens are dropped.
 
 \item \textbf{\senlabel}\citep{lin_emnlp_2019}: Cross-domain knowledge transfer based approach. A model trained on MSMARCO and TREC microblog data is used to obtain query-sentence score in a document. Finally, it aggregates document level score and top-k sentence scores (evaluated by a transfer model to compute the final document relevance score with respect to a query.
\end{enumerate}

We exclude other recently proposed approaches like  CEDR~\citep{macavaney2019contextualized:bertir:mpi} that focus on architectural engineering using BERT. Such methods are complementary to our study and can of course benefit from our analysis.

\mpara{Training details:} We train and validate using consistent and common experimental design.
The pairwise neural models are trained for a fixed number of iterations using pairwise max-margin loss and the MAP score is computed over the validation set to choose the best model. 
On the other hand, \clam{} are trained on top of BERT using binary cross-entropy loss and finally passage scores are aggregated to calculate the document level score. 
In our experiments, for fair comparisons, we use the hyper-parameters commonly used in the earlier works, i.e., sequence length of $512$, learning rate of $1e-5$, and a batch size of $16$. Learning rate is chosen based on the performance on validation set. We tried it over $1e-5$, $2e-5$, and $3e-5$ but did not observe any significant variations. We chose learning rate $1e-5$. The results are dependent on the version of Pytorch and transformer models. In this paper, we used Pytorch and transformer versions 1.7.1 and 4.10.2 respectively.

\mpara{Metrics:} We measure the effectiveness of the ranking baselines using three standard metrics -- MAP, P@20, nDCG@20~\cite{jarvelin_inf_2002}).

\mpara{Avoiding Data Leakage.}
Note that the \textit{teacher}, i.e.,  the \qamodel{}, is trained on MS-MARCO passage dataset and the queries are same (highly overlapping) as \trecdl{} dataset. Hence, we do not apply \qamodel{} over any of the test sets to avoid the data-leakage. 
Specially, this will lead to potential data leak issue for \trecdl{} dataset. However, \qalabel{} does not have this issue because training, validation, and test query sets for \trecdl{} are disjoint.

\subsection{Datasets}

We consider following four diverse TREC datasets from with varying degrees of label properties.

\begin{enumerate}
 \item \textbf{\robust}: We have 249 queries with their description and narratives. Along with queries, we also have a 528K document collection. We retrieve the top 1000 documents for each query using QLM~\citep{strohman_indri_2005}.
 
 \item \textbf{\trecdl}: The \trecdl{} document ranking dataset is divided into training, development, and test set. The training set contains around 367K queries and test set contains 200 queries. We randomly select 2000 queries to build the training set. For each of these queries, the top 100 documents are retrieved using QLM\footnote{\url{https://microsoft.github.io/msmarco/TREC-Deep-Learning-2019.html}}.
 
 \item \textbf{\core}: The \core{} contains 50 queries with sub-topics and descriptions. Queries are accompanied by a 1.8M document collection. We retrieve the top 1000 documents for each query using QLM.
 
 \item \textbf{\clueweb}: We consider the \clueweb{} dataset shared by~\citet{dai_sigir_2019}. The dataset contains 200 queries distributed uniformly in five folds and top 100 documents for each query is retrieved using QLM.
\end{enumerate}

\begin{figure}[tb]
\center
\subfloat[{\bf Query Length}]{\includegraphics[width=0.5\textwidth]{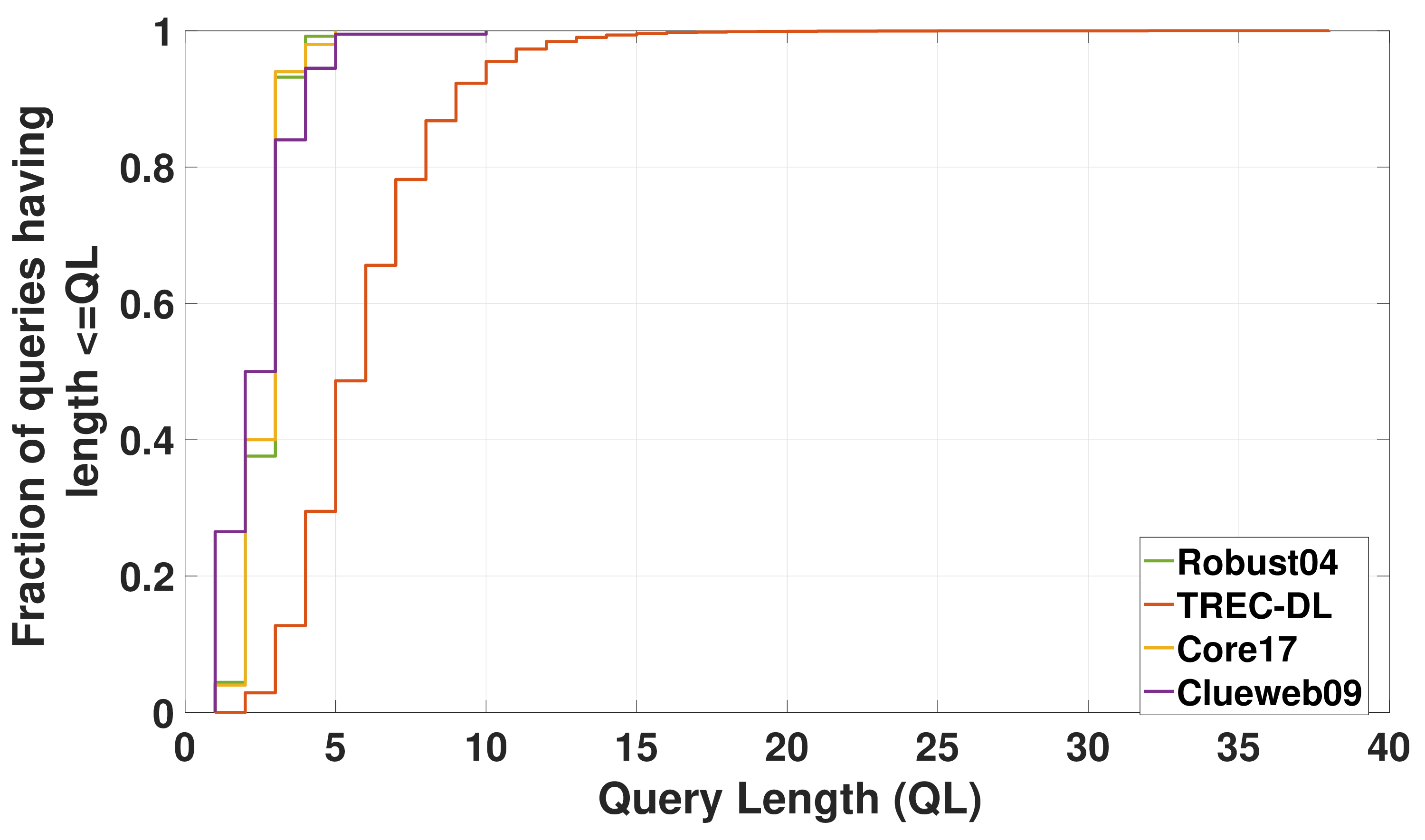}}
\hfil
\subfloat[{\bf Document Passages}]{\includegraphics[width=0.5\textwidth]{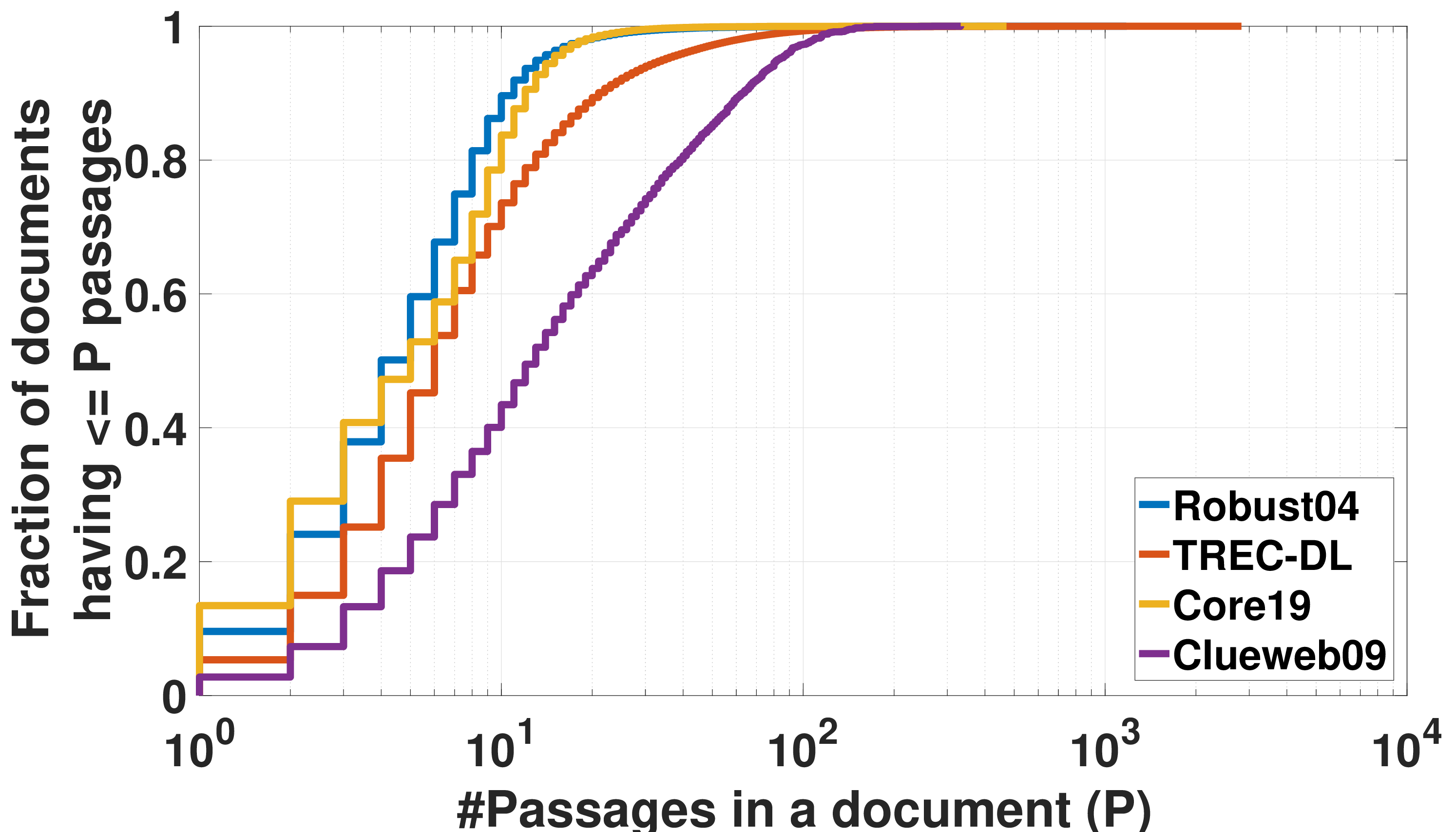}}
\caption{{CDF of query length and passages of a document.$x-axis$ is in log-scale.}}
\label{fig:query_passage_dist}
\end{figure}

The dataset details are available at \url{https://github.com/krudra/IRJ_distant_supervision_adhoc_ranking}.
Standard ad-hoc document retrieval datasets reveal quite different trend than the \trecdl{} that is curated from query logs. The queries in \trecdl{} are well specified. The documents in \trecdl{} and \clueweb{} are longer. On the other hand, news corpus  such as \robust{} and \core{} are relatively short in nature. Table~\ref{tab:dataset_query_passage_stat} provides specifications about different datasets. Figure~\ref{fig:query_passage_dist} shows the distribution of query length and the number of passages in a document.

\begin{table}[tb]
\centering
\begin{tabular}{lcc} 

\toprule
                  & \textbf{Average Query Length} & \textbf{Mean Passages}  \\ 
\midrule

\textbf{\robust} & {2.65} & {5.87} \\
\textbf{\trecdl} & {5.95} & {10.87} \\
\textbf{\core} & {2.64} & {6.26} \\
\textbf{\clueweb} & {2.47} & {23.96} \\
\bottomrule

\end{tabular}
\caption{{Statistics about the datasets. Average query length and number of passages in documents.}}
\label{tab:dataset_query_passage_stat}
\end{table}

We conduct our experiments on Nvidia 32GB V100 machine using PyTorch version 1.5.0 and evaluate baselines and our proposed models on four datasets. We have used BERT from the transformer repository (2.10.0) of Huggingface\footnote{\url{https://huggingface.co/transformers/}}. We have used deterministic version of BERT and taken a fixed seed 123 to remove the external influencing factors and make the result consistent across models.
For~\robust, \core{}, and \clueweb{} we conduct 5 fold cross-validation to minimize overfitting due to the  limited number of queries in the collection. Topics are randomly split into 5 folds and the model parameters are tuned on 4-of-5 folds. The retrieval performance is evaluated on the final fold in each case using the optimal parameters. This process is repeated five times, once for each fold. For \trecdl{}, we have 200 queries for the test set. \textit{For \clueweb, we directly take the folds from prior study~\citep{dai_sigir_2019}. \trecdl{} is also evaluated over standard test set. The folds for \robust{} and \core{} will be shared for reproducibility.}

\subsection{Results}

We elaborate the performance of \qalabel{} in this section.

\mpara{How effective is the passage level transfer for document retrieval?}

We start with comparing the ranking performance of \qalabel{} against other baselines in Table~\ref{tab:model_eval}. 

First, inline with previous works, we observe that the contextual rankers outperform other non-contextual rankers convincingly for most of the datasets.
\textbf{Among the contextual models, \senlabel{} and our approach outperform \doclabel{}.}
The improvements of \qalabel{} over \doclabel{} (nDCG20) are {\it statistically significant} with p-scores 0.002, 0.014 for \robust{} and \trecdl{} as per paired t-test ($\alpha$ = 0.05) with Bonferroni correction~\citep{paired_significance_test}. 
\senlabel{} obtains statistically significant improvement over \doclabel{} for \robust{}, \trecdl{}, and \core. 
However, the improvements of \senlabel{} over \qalabel{} are {\em not statistically significant for all the datasets.} 
As we show later that though the ranking performance obtained by \senlabel{} is competitive with our approach, their inference phase is computationally heavy (sometimes infeasible for large web collections) due to when evaluating long documents~(Section~\ref{sec:analysis}).

\begin{table}[tb]
\centering
\begin{tabular}{lcccccc}

\hline
\multicolumn{1}{l}{\multirow{2}{*}{\textbf{METHOD}}} & \multicolumn{3}{c}{\textbf{\robust}}                                                                        & \multicolumn{3}{c}{\textbf{\trecdl}}                                                                         \\ 
\cmidrule(lr){2-4}
\cmidrule(lr){5-7}
\multicolumn{1}{l}{}                                 & \multicolumn{1}{c}{\textbf{MAP}} & \multicolumn{1}{c}{\textbf{nDCG20}} & \multicolumn{1}{c}{\textbf{P@20}} & \multicolumn{1}{c}{\textbf{MAP}} & \multicolumn{1}{c}{\textbf{nDCG20}} & \multicolumn{1}{c}{\textbf{P@20}} \\ 
\midrule
\textbf{QL Model}       & 0.240 & $0.403^*$ & 0.347 & 0.237 & $0.487^*$ & 0.495    \\
\bf \pdrmm{}       & 0.263 & $0.445^*$ & 0.374 & 0.241 & $0.517^*$ & 0.508    \\
\textbf{\doclabel{}}    & 0.249 & $0.423^*$ & 0.363 & 0.258 & $0.557^*$ & 0.568    \\
\textbf{\bertcls{}}    & { 0.276} & {\bf 0.474} & {\bf 0.414} & { 0.246} & {$0.568^*$} & { 0.579}    \\
\textbf{\senlabel{}}    & {\bf 0.289} & {\bf 0.476} & {\bf 0.409} & {\bf 0.267} & {\bf 0.595} & {\bf 0.586}    \\
\textbf{\qalabel{}}     &  {\bf 0.294} & {\bf 0.471} & {\bf 0.406} & {\bf 0.269} & {\bf 0.603} & {\bf 0.602}    \\
\bottomrule
\bottomrule
\multicolumn{1}{l}{\multirow{2}{*}{\textbf{METHOD}}} & \multicolumn{3}{c}{\textbf{\core}}                                                                          & \multicolumn{3}{c}{\textbf{\clueweb}}                                                                       \\ 
\cmidrule(lr){2-4}
\cmidrule(lr){5-7}
\multicolumn{1}{l}{}                                 & \multicolumn{1}{c}{\textbf{MAP}} & \multicolumn{1}{c}{\textbf{nDCG20}} & \multicolumn{1}{c}{\textbf{P@20}} & \multicolumn{1}{c}{\textbf{MAP}} & \multicolumn{1}{c}{\textbf{nDCG20}} & \multicolumn{1}{c}{\textbf{P@20}} \\ 
\midrule
\textbf{QL Model}      & 0.203 & $0.395^*$ & 0.474 & 0.165 & $0.277^*$ & 0.331    \\
\textbf{\pdrmm{}}      & 0.215 & 0.418 & 0.497 & 0.169 & $0.285^*$ & 0.336    \\
\textbf{\doclabel{}}   & 0.239 & 0.445 & 0.514 & 0.177 & $0.309^*$ & 0.355    \\
\textbf{\bertcls{}}   & 0.242 & $0.449^*$ & 0.549 & 0.183 & $0.313^*$ & 0.354    \\
\textbf{\senlabel{}}   & {\bf 0.258} & {\bf 0.476} & {\bf 0.571} & { 0.184} & {$0.314^*$} & { 0.366}    \\
\textbf{\qalabel{}}    & {0.239} & {0.458} & { 0.539} & {\bf 0.193} & {\bf 0.341} & {\bf 0.383}    \\
\bottomrule
\end{tabular}
\caption{{Retrieval performance of baselines, and \qalabel{} method. We report the result of the best aggregation method for \qalabel{} and \doclabel{}. For \doclabel{} MaxP and DecaySumP show best result for (\robust, \core, \trecdl), and \clueweb{} respectively. For \qalabel{} MaxP and DecaySumP show best result for (\robust, \trecdl), and (\core, \clueweb) respectively. $*$ implies \qalabel{} is statistically significantly better at 95\% significance level, than the corresponding baseline method.}}
\label{tab:model_eval}
\end{table}

\mpara{Threats to Validity.} We note that there are differences between the ranking performance of baselines as measured by us and in the original paper of the authors and that can be attributed to experimental design choices for better comparison. The difference in \doclabel{} is mainly due to the differences in passage chunking setup. The folds are different in case of \robust. The implementation set up is different than the original paper. We use full token length (512) of BERT instead of 256 token length and implement the code in Pytorch using a deterministic version of BERT\footnote{\url{https://huggingface.co/docs/transformers/model_doc/bert}}. The performance of the model depends on the version of BERT model. Hence, we set a specific seed value to make the results reproducible. We train the entire setup using Pytorch version $1.7.1$ and Transformer version $4.10.2$. 
We attribute the performance difference of \senlabel{} occurred for the following three design choices: (i).~they selected BERT-LARGE as their fine-tuning model. 
However, we choose BERT-BASE to make a fair performance comparison among the three BERT based models. (ii).~they used MSMARCO+MICROBLOG based transfer model to learn the relevance of query-sentence pairs. We only select MSMARCO as the transfer model to keep consistency between \qalabel{} and \senlabel{}. (iii).~they used BM25+RM3 instead of QL to retrieve initial document set. The objective of this paper is to compare the effectiveness-efficiency trade-off of BERT based reranking models and their variation based on granularity of documents (passage/sentence), label transfer etc. Hence, we try to keep external influencing factors almost same across different models. However, we believe that considering BERT-LARGE and MSMARCO+MICROBLOG based transfer model will not change the trend  significantly.

\section{Analysis}
\label{sec:analysis}

We present our effectiveness-efficiency related findings based on the research questions formulated in the Section~\ref{sec:intro}.

\begin{figure}[tb]
\center
\subfloat[{\bf Small Training sets}]{\includegraphics[width=0.5\textwidth]{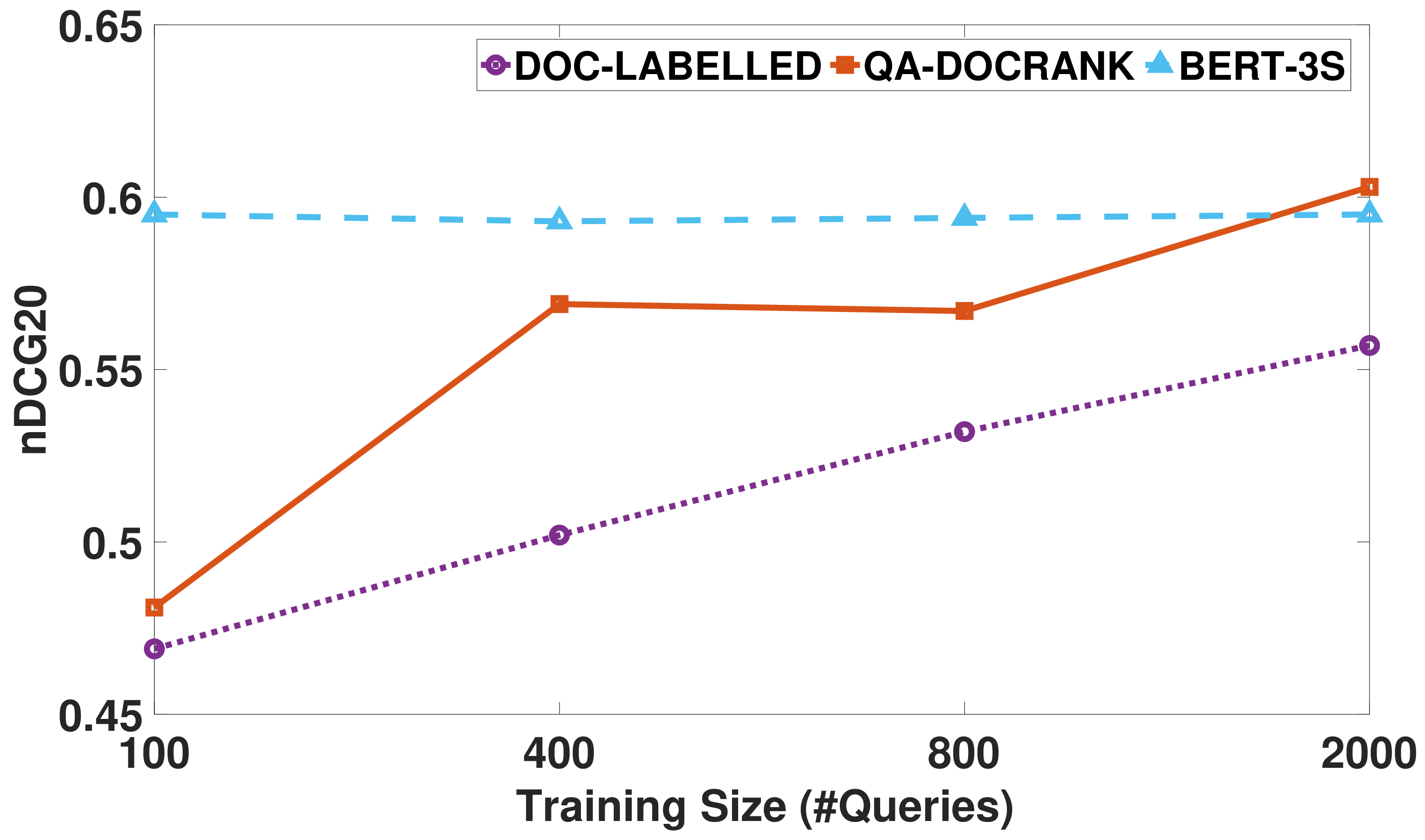}}
\hfil
\subfloat[{\bf Larger Training Sets}]{\includegraphics[width=0.5\textwidth]{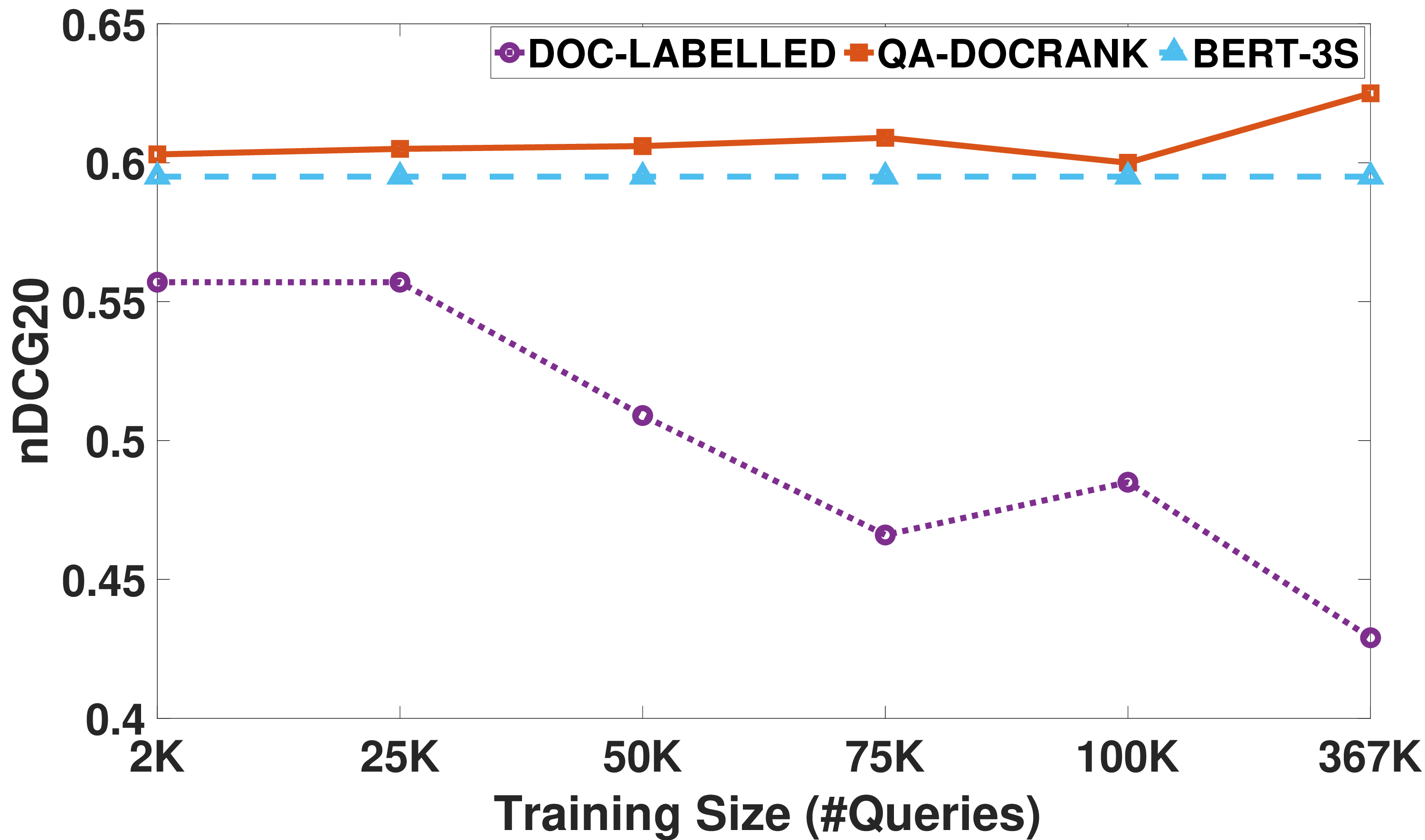}}
\caption{{Effect of training dataset size on ranking performance. (a) Small training set, (b) larger training sets.}}
\label{fig:query_vs_performance}
\end{figure}

\subsection{How do larger training datasets impact training of contextual ranking models?}

We now measure the effectiveness of ranking models for different training data sizes -- measured by number of queries.
We consider three contextual models and best performing pairwise neural model (\pdrmm). 

First we look at performance of rankers in small data regime reported in Figure~\ref{fig:query_vs_performance} (a). 
We randomly select $100, 400, 800$ queries from the training set of \trecdl{} to train the models. 
We observe that for smaller datasets, \qalabel{} suffers due to its high selectivity that results in small number of training instances (see nDCG@20 for 100 queries).
However, the performance monotonically increases with the number of queries and is already equivalent to \senlabel{} for 1000 queries. \textit{Note that, performance of \senlabel{} is constant because interpolation parameter $\alpha$ is learned on validation set and training set has no impact on it.}

Next, we analyze the effect of using larger datasets on passage-level BERT models. 
Figure~\ref{fig:query_vs_performance}(b) presents the ranking performance with increasing training data from 2K queries to 100K queries from the \trecdl{} dataset. All results are reported over the same test set of 2K queries. 
\textbf{We find that the performance of \doclabel{} is significantly affected with increasing number of training queries.}
We attribute this to the longer documents in \trecdl{}. 
Specifically, the average document length of \trecdl{} is almost twice that of \robust{} and \core. 
\textit{This means that longer relevant Web documents tend to contain more irrelevant passages and document to passage label transfer is susceptible to higher label noise.}
This, along with results from the passage generation experiment, clearly establishes the negative impact of label noise introduced by document to passage label transfer. 
On the other hand, \qalabel{} is effectively able to filter out noise due to its judicious label selection strategy and is unaffected by increasing training size.

\begin{table}[tb]
\centering
\begin{tabular}{lcccccc} 

\toprule
                  & \textbf{FirstP} & \textbf{MaxP} & \textbf{SumP} & \textbf{AvgP} & \textbf{DecaySumP} & \textbf{DecayAvgP} \\ 
\midrule

\textbf{\robust} & {0.420} & {\bf 0.471} & {0.418} & {0.365} & {0.444} &  {0.298} \\
\textbf{\trecdl} & {0.581} & {\bf 0.603} & {0.520} &  {0.525} & {0.568}  & {0.449} \\
\textbf{\core} & {0.391} & {0.408} & {0.437} & {0.365} & {\bf 0.458} & {0.273} \\
\textbf{\clueweb} & {0.313} & {0.308} & {0.313} & {0.285} & {\bf 0.341}  & {0.218} \\
\bottomrule

\end{tabular}
\caption{{Retrieval performance (nDCG20 score) of \qalabel{} over different passage score aggregation set-ups.}}
\label{tab:aggregation_result}
\end{table}

\subsection{What is the role of passage aggregation strategy on overall performance?}
It is evident from Table~\ref{tab:model_eval} that the performance of the models over different datasets very much sensitive to the aggregation strategies. For \robust{} and \trecdl, the maximum score of a passage turns out to be a good measure for the entire document. On the other hand, position decay weighted summation performs better than other aggregation strategies for \core{} and \clueweb. Table~\ref{tab:aggregation_result} shows the performance of \qalabel{} over different datasets for different aggregation strategies. The results suggest that the ranking strategy should dynamically determine the aggregation strategy and end-to-end setup shows promising results due to the ability of this data specific adaptation~\citet{li2020parade}.

\begin{table}[tb]
\centering
\begin{tabular}{lcccc} 

\toprule
\multirow{2}{*}{} & \multicolumn{2}{c}{\qalabel}  & \multicolumn{2}{c}{\doclabel}                      \\
\cmidrule(lr){2-3}
\cmidrule(lr){4-5}
                  & \textbf{$P\_100$} & \textbf{$P\_150\_75$} & \textbf{$P\_100$} & \textbf{$P\_150\_75$} \\ 
\midrule

\textbf{\robust} & {\bf 0.471} & {0.462} & {0.423} & {0.436} \\
\textbf{\trecdl} & {\bf 0.627} & {0.555} & {0.429} & {0.434} \\
\textbf{\core} & {\bf 0.458} & {\bf 0.456} & {0.445} & {0.437} \\
\textbf{\clueweb} & {\bf 0.341} & {0.316} & {0.309} & {0.289} \\
\bottomrule

\end{tabular}
\caption{{Retrieval performance (nDCG20 score) of \qalabel{} and \doclabel{} over two different passage generation set-ups.}}
\label{tab:passage_variation_result}
\end{table}


\subsection{How robust are the models to passage generation?}
In the previous part, we observe that performance of \doclabel{} drops with the increase of training query size. Here, our objective is to check the variation in the performance with the passage sizes.
\doclabel{} splits documents into passages of length $150$ words with an overlap of $75$ words between consecutive passages and consider $30$ passages (first,last, and random $28$). However, we generate non-overlapping passages of length 100 following the approach proposed in~\citet{fan_sigir_2018} since we consider it a better way for passage splitting. 
To verify the superiority of our proposed approach, we also apply \qalabel{} and \doclabel{} to the passages generated using the approach mentioned in~\citet{dai_sigir_2019}\footnote{We are grateful to the authors for sharing the data.}.

Table~\ref{tab:passage_variation_result} shows the nDCG scores of both the methods under different passage generation setups over all the four datsets.
Note that, for \doclabel{} in \trecdl{} we have used the entire training query set (367K) instead of 2000 queries; hence, the nDCG score in Table~\ref{tab:passage_variation_result} under $P_{100}$ column ($0.429$) is different from the value reported in Table~\ref{tab:model_eval} ($0.541$).
\qalabel{} is robust to passage generation setup. However, \doclabel{} is very much sensitive to the passages. We find that document to passage level label transfer introduces significant noise in the training phase.

\begin{figure}[tb]
\center
\subfloat[{\bf \core}]{\includegraphics[width=0.5\textwidth]{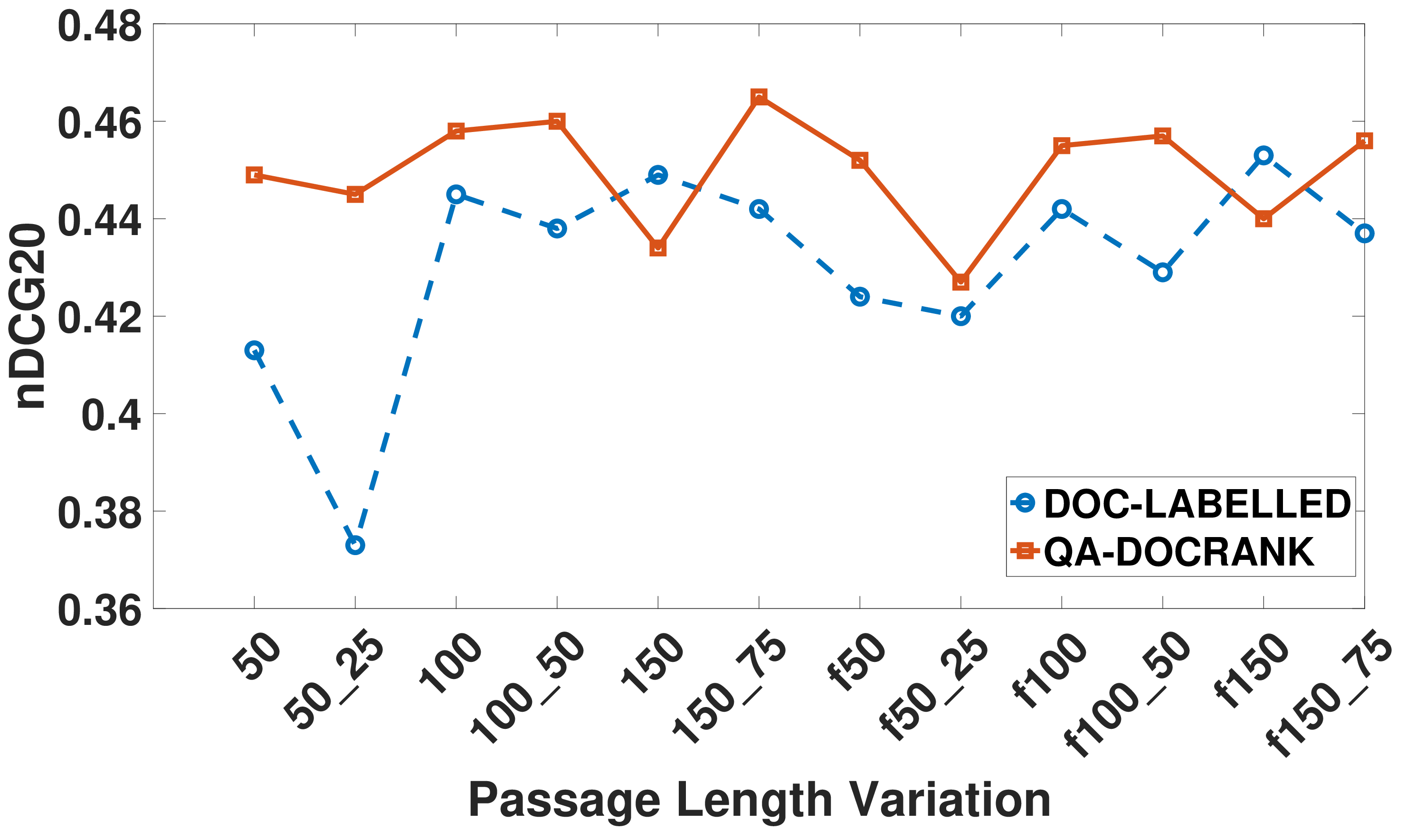}}
\hfil
\subfloat[{\bf \robust}]{\includegraphics[width=0.5\textwidth]{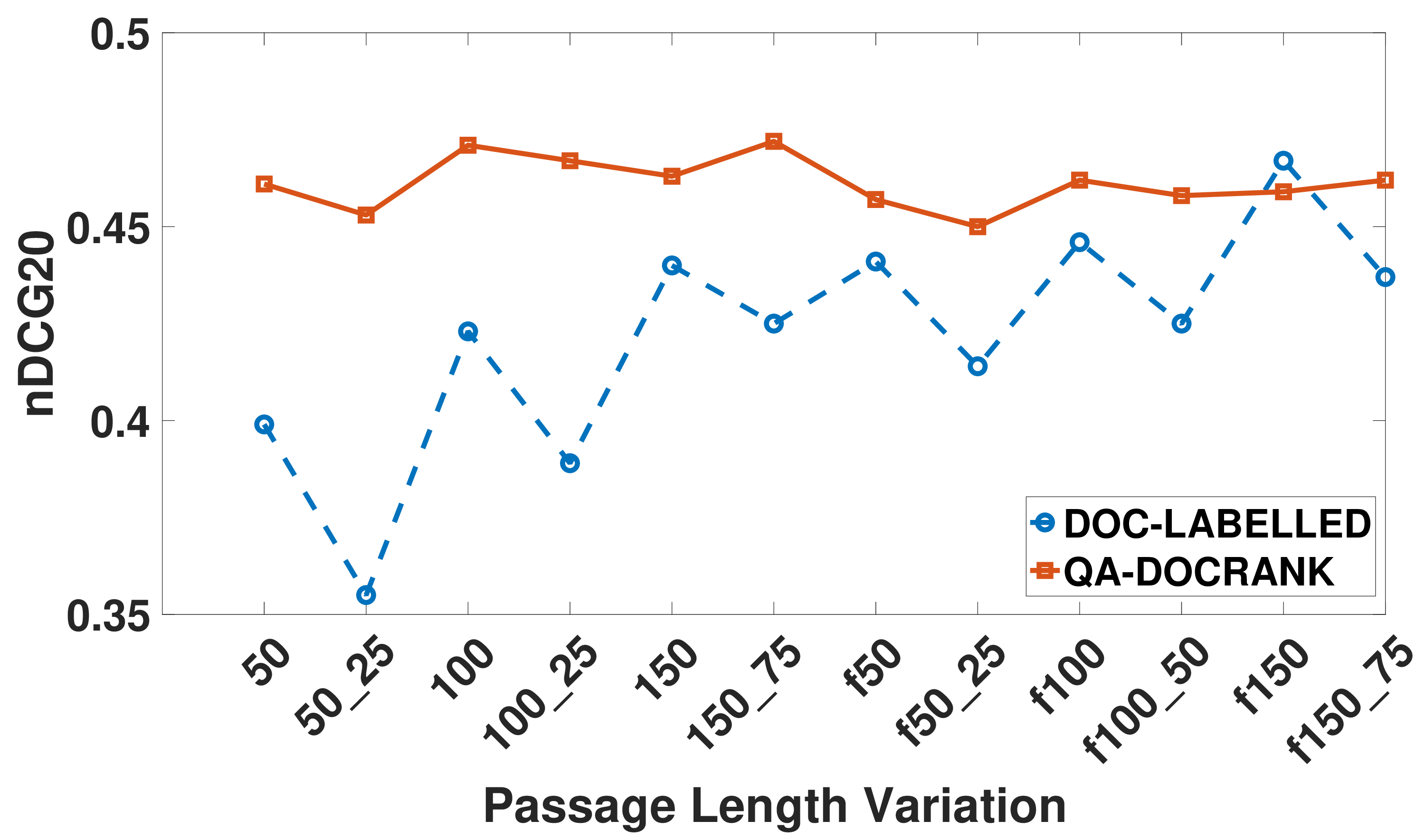}}
\caption{{Effect of passage granularities on the performance of. (a) \core, (b) \robust.}}
\label{fig:passage_vs_performance}
\end{figure}

\noindent\textbf{Variation in performance based on passage size: }
Both \doclabel{} and \qalabel{} are dependent on the input size limit enforced by BERT models. In this experiment we want to first evaluate if different granularities of partitioning documents into passages affect ranking performance. 
We study the ranking performance based on varying ---~\textit{passage length},~\textit{overlap between two consecutive passages}, and~\textit{number of passages}. 
This entails four scenarios -- (i).~${X}:$ documents are split into passages of length $X$ and all the passages are considered, (ii).~${X\_Y}:$ similar to case (i) but there is an overlap of $Y$ words between two consecutive passages, (iii).~$f{X}:$ passages are of length $X$. Following the approach in~\cite{dai_sigir_2019}, the first, last, and randomly 28 other passages are chosen instead of all the passages, and (iv).~$f{X\_Y}:$ similar to case (iii) with overlap of $Y$ words. 
Figure~\ref{fig:passage_vs_performance} reports the performance under the above mentioned four scenarios for different $X$ and $Y$.
The major takeaway from this experiment is that \doclabel{} is sensitive to passage generation while \qalabel{} is robust.
We observe that the performance of \doclabel{} improves considerably when fixed number of passages are considered. 
This is the first evidence that direct assignment of labels to all constituent passages is wasteful and leads to label noise.
A fixed number of passages implicitly controls label noise.

\begin{figure}[tb]
\centering
\includegraphics[width=0.8\textwidth]{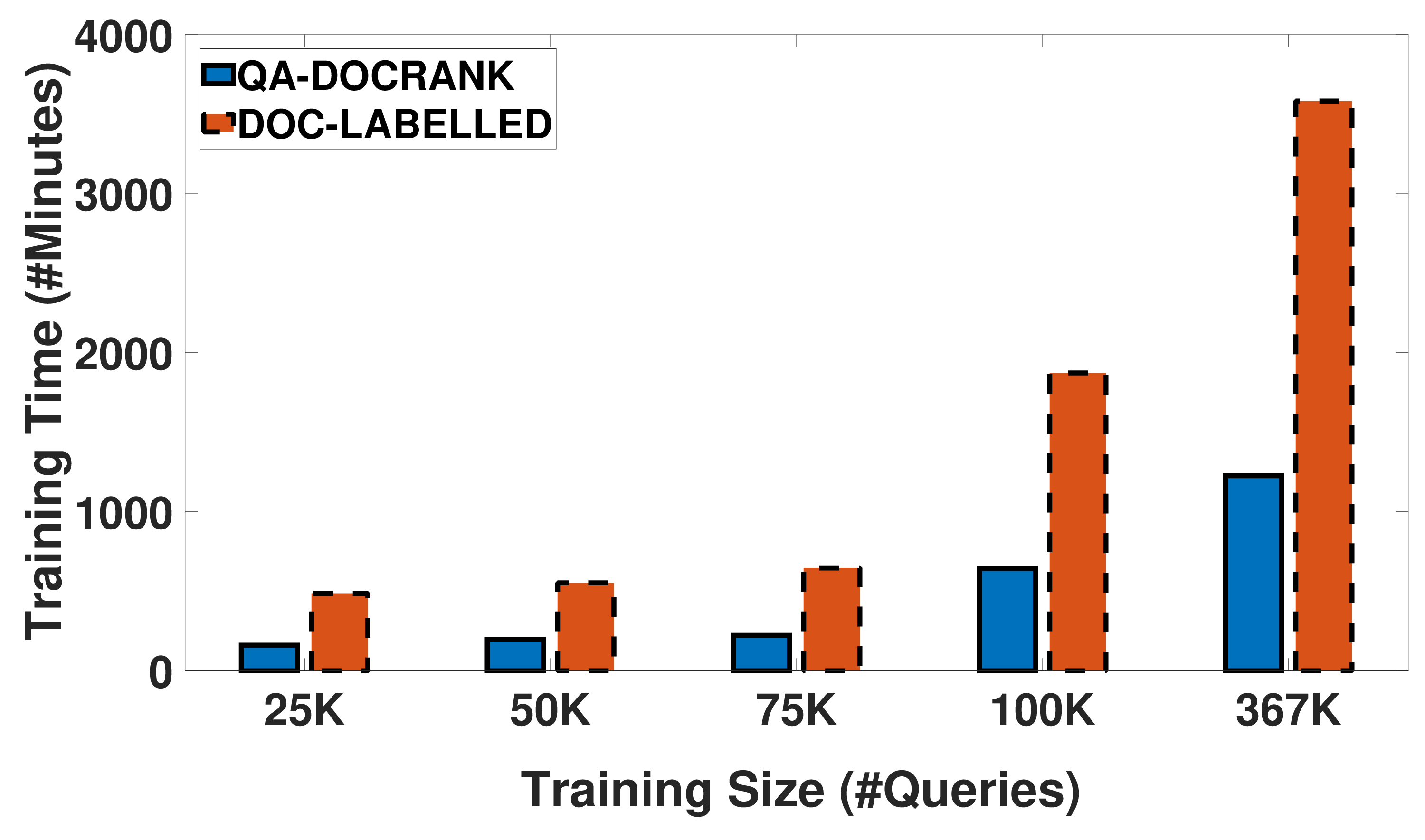}
\caption{{Effect of training dataset size on training time for \qalabel{} and \doclabel{}.}}
\label{fig:training_time_with_queries}
\end{figure}


\subsection{How efficient is it to {\em train} transfer-based BERT retrieval models?} 

In this experiment we measure the wall-clock times (in minutes) for training BERT-based models.
Firstly, we note that \senlabel{} does not involve any fine-tuning in the training phase and is not included in the experiment.
On the other hand, \qalabel{}, and \doclabel{} require fine-tuning over the task specific dataset. 
For \qalabel{}, we also consider the time taken by \qamodel{} to find out relevant training passages. However, training/fine-tuning time of \qamodel{} is not considered following the same protocol as \senlabel{}.
As expected, we observe from Figure~\ref{fig:training_time_with_queries} that the \textbf{training time of \doclabel{} is $2.5-3$ times higher than \qalabel{}}. 
This is a direct impact of selective labelling of passages in \qalabel{} that results in far fewer training instances in comparison to \doclabel{} that indiscriminately transfers labels to all the passages. 
Specifically, the training size of \qalabel{} is around 7-8\% of \doclabel{} approach. 
This further strengthens the hypothesis that selective passage transfer not only helps in retrieval effectiveness but has a direct impact on training efficiency by being sample efficient.

\begin{figure}[tb]
\centering
\includegraphics[width=0.8\textwidth]{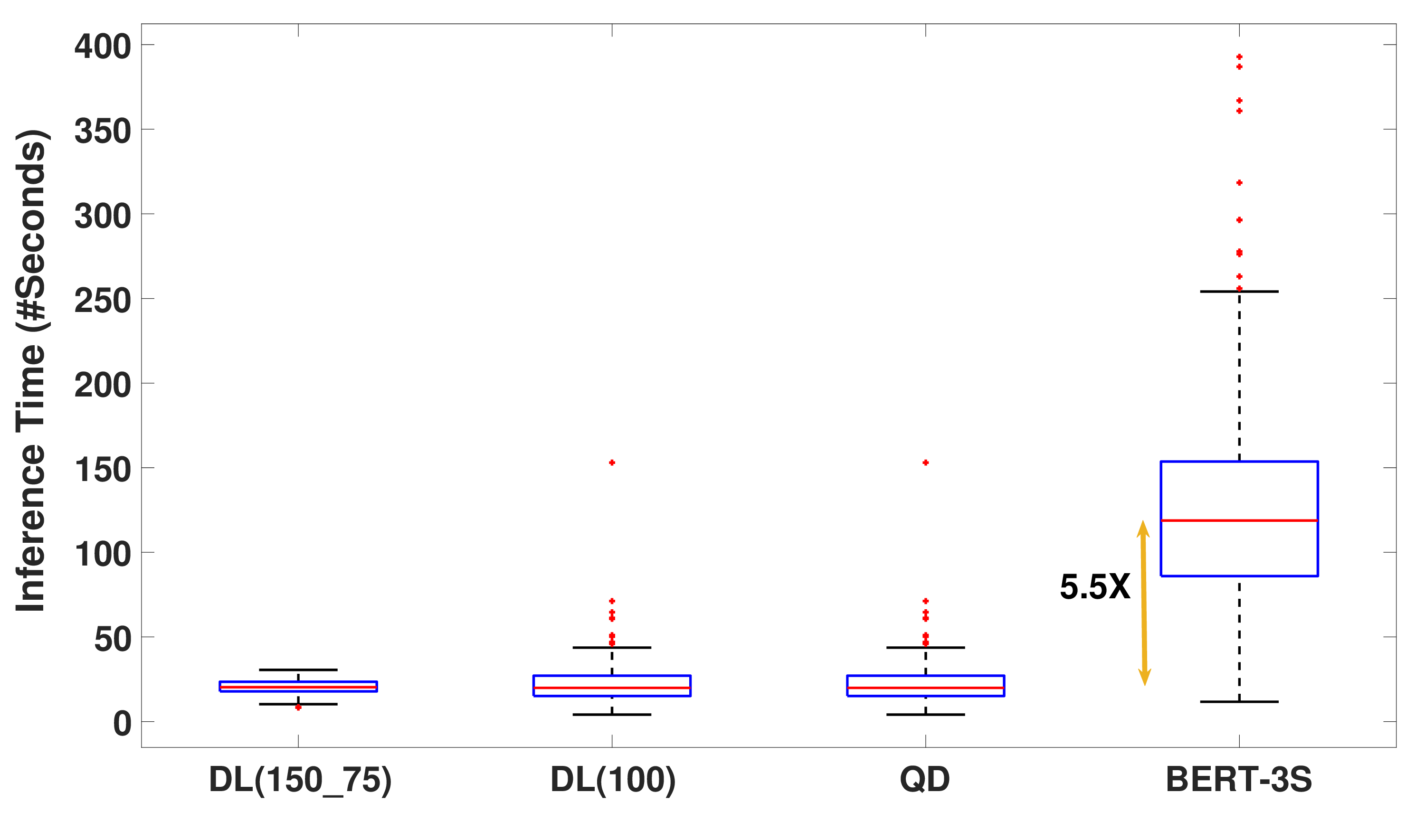}
\caption{{Comparison of inference times for BERT based models (\doclabel($DL(150\_75)$), \doclabel($DL(100)$), \qalabel, \senlabel). $DL(150\_75)$ consists of passage length 150 words with an overlap of 75 words between consecutive passages. Median inference time for \senlabel{} is 5.5 times higher than \qalabel{} and \doclabel{}.}}
\label{fig:inference_time_with_queries}
\end{figure}

\subsection{How efficient is {\em inference} for transfer-based BERT retrieval models?} 

Figure~\ref{fig:inference_time_with_queries} reports the variation in wall-clock times for query processing or inference of \qalabel{}, \doclabel{}, and \senlabel{} over the 200 test queries of \trecdl. We only measure the inference time of the model and do not include the preprocessing times such as tokenization of query and documents, batching etc.
For \doclabel{}, we test both the variations of passage chunking. In $DL(150\_75)$, passages are created as proposed in~\citet{dai_sigir_2019} i.e., each passage is 150 words long with an overlap of 75 words between consecutive passages and 30 passages are selected. On the other hand, $DL(100)$, passages are 100 words long and there is no overlap between passages.  
As expected, we observe that the \textbf{average query processing time for \senlabel{} is much larger than \qalabel{} due to the sentence-level scoring} -- around 5.5 times higher than \qalabel{} and \doclabel{} approach. There is no significant difference in mean and median between \doclabel{} and \qalabel. The average time taken by $DL(150\_75)$ and $DL(100)$ is almost same. However, the standard deviation of $DL(100)$ is 3 times higher than $DL(150\_75)$ because later version restricts to 30 passages per document.
\senlabel{} also has a large standard deviation $4\times$ higher than \qalabel{} with some queries with long result documents taking $400$ seconds to process.
These results reflect on an yet to be resolved open question in terms of efficient inference of BERT-based models.
\textit{We did not consider any parallel optimization techniques in the score computation process for any of the methods. We believe that each method will get similar kind of improvement in the running time (i.e., inferring scores of passages/sentences).}

\begin{table}[tb]
\centering
\begin{tabular}{lcccc} 

\toprule
\diagbox{{\bf Train}}{{\bf Test}} & \textbf{\robust} & \textbf{\trecdl} & \textbf{\core} & \textbf{\clueweb} \\
\midrule

\textbf{\robust} & {0.471} & {0.519(0.528)} & {0.435} & {0.245(0.301)} \\
\textbf{\trecdl} & {\bf 0.459(0.471)} & {\bf 0.627} & {0.410(0.443)} & {0.291 (0.312)} \\
\textbf{\core} & {\bf 0.454(0.464)} & {0.533} & {\bf 0.458} & { 0.314} \\
\textbf{\clueweb} & 0.443 & {0.599} & {0.424} & {\bf 0.341} \\
\bottomrule

\end{tabular}
\caption{{Cross-Domain Retrieval performance (nDCG20 score) of fine-tuned \qalabel{} with the aggregation strategy same as training dataset. Values in the bracket presents the best score achieved through some aggregation strategy that is different from the training set.}}
\label{tab:cross_domain_result}
\end{table}

\subsection{How well does a model fine-tuned on one document transfer to another collection?}
\label{sec:zero-shot}
In the last section, we check the efficacy of {\it transfer learning} from \qamodel{} to document ranking (\qalabel{} (passage-level), \senlabel{}(sentence-level)). Interestingly, it performs quite well even without any fine-tuning in some cases (\senlabel{}). Here, we verify effectiveness of document-specific fine-tuned models. We fine-tune our \qalabel{} model over a dataset and test it over the other ones. 
It is interesting to note that in some cases cross-document testing gives almost similar ranking like the models trained on the same document.
Table~\ref{tab:cross_domain_result} shows the results on cross domain inference. In cross-domain setup, we don't have any clues about the test data set; hence, we have to rely on the aggregation strategy that works best for the training dataset. Table~\ref{tab:cross_domain_result} reports both results i.e., results achieved on the test data based on the aggregation strategy of the training data and the best score achieved through another aggregation strategy(in brackets).
Specifically \qalabel{} shows comparable performance to the in-domain testing but this depends on the training dataset. For example, \core{} achieves best ranking for \robust{} and \clueweb{} whereas \clueweb{} performs well for \trecdl. \robust{} and \core{} both are news corpus whereas \clueweb{} and \trecdl{} are web corpus; hence, these groups follow different information distribution. In general, news corpus contains information mostly in first couple of passages. \trecdl{} contains significantly large number of queries than other datasets; hence, training and generalization of models are quite easy for this case. \trecdl{} shows consistent performance over other datasets. 
However, performance of other models is also not significantly bad than the best performer. 
It indicates that \textbf{careful selection of training documents and passages might help in the direct application of document-specific fine-tuned models over new collection.}

\section{Conclusion}
\label{sec:conclusion}

In this paper, we illustrate the shortcomings of two transfer learning based modeling approaches \doclabel{} and \senlabel{}. The former one suffers due to label noise that degenerates its performance beyond a certain point while the inference time restricts the utility of the later one. We have combined the positive aspects of both the models and proposed an approach that optimizes both {\bf retrieval performance} and {\bf computational complexity}. We also show the robustness of this model towards document splitting schemes and its applicability in cross-domain document ranking i.e., model trained on one document set may be directly applied to another set. 

Throughout this paper, we assume that passages are disjoint in nature and treat them as separate entities during relevance prediction. In the future, it is interesting to capture the interaction among different passages to check its impact in retrieval performance. A very few passages in a document are ultimately relevant to a given query. Hence, it will be interesting to find out such denoised version of the document before retrieval and ranking task. This will also be helpful to bring interpretability into the framework. We also explore large language models to  improve query rewriting and zero-shot model performance.

\bibliographystyle{plainnat}
\bibliography{main}

\end{document}